\documentclass[aps,pre,reprint,unsortedaddress]{revtex4-1}

\usepackage{graphicx}
\usepackage{latexsym}
\usepackage{natbib}
\usepackage{float}
\usepackage[caption=false]{subfig}

\usepackage{amssymb}
\usepackage{amsmath}
\usepackage{mathptmx}

\begin{document}

\title{Fragmentation transition in a coevolving network with link-state
dynamics}

\author{A. Carro}
\email[Electronic address: ]{adrian.carro@ifisc.uib-csic.es}
\affiliation{IFISC, Instituto de F\'isica Interdisciplinar y Sistemas Complejos
(CSIC-UIB), E-07122 Palma de Mallorca, Spain}
\author{F. Vazquez}
\affiliation{IFLYSIB, Instituto de F\'isica de L\'iquidos y Sistemas
Biol\'ogicos (UNLP-CONICET), 1900 La Plata, Argentina}
\author{R. Toral}
\author{M. San Miguel}
\affiliation{IFISC, Instituto de F\'isica Interdisciplinar y Sistemas Complejos
(CSIC-UIB), E-07122 Palma de Mallorca, Spain}

\date{\today}

\begin{abstract}
We study a network model that couples the dynamics of link states with the
evolution of the network topology. The state of each link, either $A$ or $B$, is
updated according to the majority rule or zero-temperature Glauber dynamics, in
which links adopt the state of the majority of  their neighboring links in the
network. Additionally, a link that is in a local minority is rewired to a
randomly chosen node. While large systems evolving under the majority rule alone
always fall into disordered topological traps composed by frustrated links, any
amount of rewiring is able to drive the network to complete order, by relinking
frustrated links and so releasing the system from traps. However, depending on
the relative rate of the majority rule and the rewiring processes, the system
evolves towards different ordered absorbing configurations: either a
one-component network with all links in the same state or a network fragmented
in two components with opposite states. For low rewiring rates and finite size
networks there is a domain of bistability between fragmented and non-fragmented
final states. Finite size scaling indicates that fragmentation is the only
possible scenario for large systems and any nonzero rate of rewiring.
\end{abstract}

\maketitle

\section{Introduction}
\label{intro}


The emergence of collective properties in systems composed of many interacting
units has traditionally been studied in terms of some property or state
characterizing each of these individual units. In this approach, the result of
any given interaction depends on the states of the units involved and the
particular interaction rules implemented. This basic setup, initially inspired
in the realm of physics by the study of spin systems, has been also extensively
used for the analysis of social systems, where the variable assigned to each
agent can be for example an opinion state, a political alignment, a religious
belief, the competence in a given language, etc \cite{Castellano2009}. However,
there is a number of situations in which the variable of interest is a
characteristic of the interaction link instead of an intrinsic feature of each
interacting unit. This is particularly the case when studying some social
interactions such as friendship-enmity relationships, trust, communication
channel, method of salutation or the use of competing languages.


There are in the literature three main areas where a focus has been placed on
link properties and their interactions: social balance theory, community
detection and network controllability. Social balance theory \cite{Heider1946}
is the first and most established precedent. Assuming that each link or social
relationship can be positive or negative, this theory proposes that there is a
natural tendency to form balanced triads, defined as those for which the product
of the states of the three links is positive. The question of whether a balanced
global configuration is asymptotically reached for different network topologies
has been addressed by several recent studies \cite{Antal2005, Antal2006,
Radicchi2007}. Large scale data on link states associated with trust, friendship
or enmity has recently become available from on-line games and on-line
communities, providing an ideal framework to test the validity of this theory
and propose alternative interaction rules \cite{Szell2010, Leskovec2010A,
Leskovec2010B, Marvel2011}. The problem of community detection in complex
networks has been addressed in a number of recent works \cite{Traag2009,
Evans2009, Evans2010, Ahn2010, Liu2012} using a description in terms of link
properties. Identifying network communities to sets of links, instead of sets of
nodes \cite{Fortunato2010}, allows for an individual to be assigned to more than
one community, which naturally gives rise to overlapping communities, a problem
difficult to tackle from the traditional node perspective. Finally, the
controllability of networks, that is, the problem of determining the conditions
under which the dynamics of a network can be driven from any initial state to
any desired final state within finite time, has also been recently considered
from a link dynamics perspective \cite{Nepusz2012}. The aim is therefore to
identify the most influential links for determining the global state of the
network.


In this context, a simple prototype model for the dynamics of link states in a
fixed complex network has been recently introduced by J. Fern\'andez-Gracia
\textit{et al.} \cite{Fernandez-Gracia2012}. In this model, each link can be in
one of two equivalent states and the dynamics implemented is a simple majority
rule for the links, so that in each dynamical step the state of a randomly
chosen link is updated to the state of the majority of its neighboring links,
i.e., those sharing a node with it. The authors find a broad distribution of
non-trivial asymptotic configurations, including both frozen and dynamically
trapped configurations. Some of these asymptotic disordered global states have
no counterpart under traditional node dynamics in the same topologies, and those
which have a nodal counterpart appear with a significantly increased probability
under link dynamics. These results can be qualitatively understood in terms of
the implicit topological difference between running a given dynamics on the
nodes and on the links of the same network. Indeed, one can define a
node-equivalent graph by mapping the links of the original network to nodes of
a new one, known as line-graph \cite{Rooij1965, Krawczyk2011}, where nodes are
connected if the corresponding links share a node in the original network.
Line-graphs are characterized by a higher connectivity \cite{Chartrand1969} and
a larger number of cliques \cite{Manka-Krason2010}, which results in more
topological traps and therefore a wider range of possible disordered asymptotic
configurations.


In this article, we study a coevolution model that couples this majority rule
dynamics of link states with the evolution of the network topology. The study
of coevolving dynamics and network topologies has received much attention
recently \cite{Gross2008, Herrera2011, Sayama2013}, particularly in the context
of social systems and always from a node states perspective. In the most common
coupling scheme, node states are updated according to their neighbors' states
while links between nodes are rewired taking into account the states of these
nodes. This coupled evolution generally leads to the existence of a
fragmentation transition: for a certain relation between the time scales of
both processes, the network breaks into disconnected components. A large number
of dynamics and rewiring rules have been studied \cite{Zimmermann2001,
Zimmermann2004, Holme2006, Vazquez2007, Vazquez2008a, Mandra2009, Demirel2014}.
As in \cite{Fernandez-Gracia2012}, we consider a link-state dynamics where each
link can be in one of two equivalent states and they are updated according to
the majority rule or zero-temperature Glauber dynamics \cite{Haggstrom2002,
Castellano2005, Castellano2006, Baek2012}, in which links adopt the state of the
majority of their neighboring links in the network. Additionally, we define a
rewiring mechanism inspired by the case of competing languages. In the context
of language competition dynamics, language has been so far modeled as an
individual property \cite{Abrams2003, Castello2006, Patriarca2012,
Castello2013}. However, the use of a language, as opposed to its knowledge or
the preference for it, can be more clearly described as a characteristic of the
interaction between two individuals than a attribute of these individuals. In
this way, different degrees of bilingualism arise naturally as a characteristic
of those individuals who hold at least one conversation in each of the two
possible languages. The rewiring mechanism implemented captures the fact that,
when an agent is uncomfortable with the language of a given interaction, she can
both try to change that language or simply stop this interaction and start a new
one in her preferred language. We find that depending on the relative rate of
the majority rule and the rewiring processes, the system evolves towards
different absorbing configurations: either a one-component network with all
links in the same state or a network fragmented in two components with opposite
states. It turns out that large systems evolving under the majority rule alone
always fall into topological traps which prevent total ordering, as shown in
\cite{Fernandez-Gracia2012}. Interestingly, even a very small amount of
rewiring is enough to slowly drive the network to complete order, understood as
the absence of common nodes between links in different state, independently of
the fragmentation or not of the network. For finite systems and low rewiring we
find a region of bistability between fragmented and non-fragmented absorbing
states. Increasing rewiring leads always to the fragmentation of the network
into two similar size components with different link states. By means of a
scaling analysis we show that the bistability region vanishes as the system size
is increased, and thus fragmentation is the only possible scenario for large
coevolving systems. We also show that a mean-field approach is able to describe
the ordering of the system and its average time of convergence to the final
ordered state for large rewiring values.

The paper is organized as follows. In section \ref{model} we define the rewiring
mechanism which is coupled with the majority rule of link states to produce a
coevolving model. We also present in this section a schematic view of the
results obtained with the majority rule alone and some quantities introduced for
its characterization. In section \ref{finalStates} we describe the final states
obtained with the coevolving model and we characterize the observed
fragmentation transition (subsection \ref{fragTrans}). In section
\ref{timeEvol} we study the time evolution of the system, including a
description of the trajectories in phase space (subsection \ref{trajectories}),
a mean-field approach for the order parameter (subsection \ref{meanField}) and
an analysis of the times of convergence to the final ordered state (subsection
\ref{convTimes}). Finally, section \ref{conclusions} contains a discussion
summary.


\section{The Model}
\label{model}

\begin{figure}[ht!]
 \centering \includegraphics[width=\columnwidth, height=!]{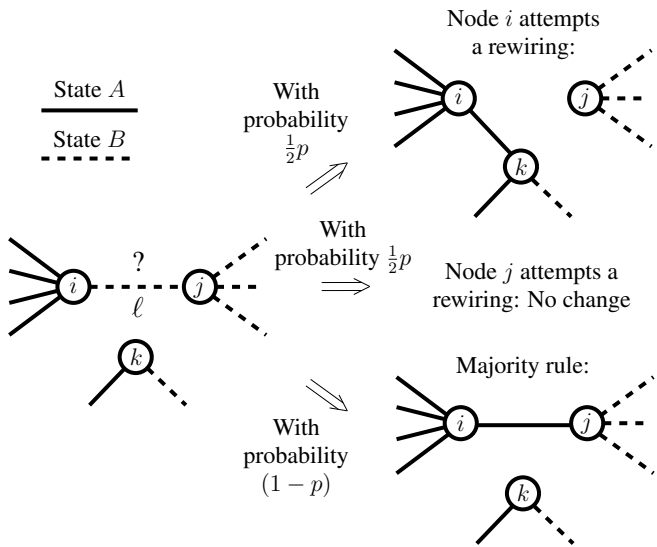}
 \caption{Schematic illustration of the dynamics for both a successful and a
 failed rewiring attempt and the application of the majority rule.}
 \label{algorithm}
\end{figure}

We consider an initially connected Erd\"os-R\'enyi random network composed by a
fixed number of nodes $N$ and with a fixed mean degree $\mu \equiv \langle k
\rangle$. The state of each link $\ell$ is characterized by a binary variable
$S_l$ which can take two equivalent or symmetrical values, for example, $A$ and
$B$. Link states are initially distributed with uniform probability. At each
time step, a link $\ell$ between nodes $i$ and $j$ is chosen at random. Then,
with probability $p$ a rewiring event is attempted (see Fig.~\ref{algorithm}
for a schematic illustration of the dynamics): one of the two nodes at the ends
of $\ell$, for example, $i$, is chosen at random and
\begin{enumerate}
\item if $S_l$ is different from the state of the majority of links attached to
  $i$, then the link $\ell$ is disconnected from the opposite end, $j$, and
  reconnected to another node, $k$, chosen at random, and also its state $S_l$
  is switched to comply with the local majority around node $i$;
\item otherwise, nothing happens.
\end{enumerate}
With the complementary probability, $1-p$, the majority rule is applied: the
chosen link, $\ell$, adopts the state of the majority of its neighboring links,
i.e., those links connected to the ends of $\ell$ (nodes $i$ and $j$). In case
of a tie, $\ell$ switches state with probability $1/2$. Finally, time is
increased by $1/N$, so that for each node, on average, the state of one of its
relationships is updated per unit time. In this manner, the time scale of the
process for each agent becomes independent of system size for constant degree
distribution.

The rewiring mechanism mimics the fact that, when a speaker is uncomfortable
with the language used in her interaction with other speaker, one of her
possibilities is to stop this relationship and start a new one in her preferred
language with any other individual. The majority rule mechanism captures the
fact that the language spoken in a given interaction tends to be that most
predominantly used by the interacting individuals, that is, the one they use
more frequently in their conversations with other people. In this way, agents
tend to avoid the cognitive cost of speaking several languages. The rewiring
probability $p$ measures the speed at which the network evolves, compared to
the propagation of link states. It is, therefore, a measure of the plasticity
of the topology. When $p$ is zero the network is static and only the majority
rule dynamics takes place (as studied in \cite{Fernandez-Gracia2012}), while in
the opposite situation, $p=1$, there is only rewiring. 

\begin{figure}[ht!]
  \subfloat[\label{frozen}]{
    \includegraphics[width=0.46\columnwidth]{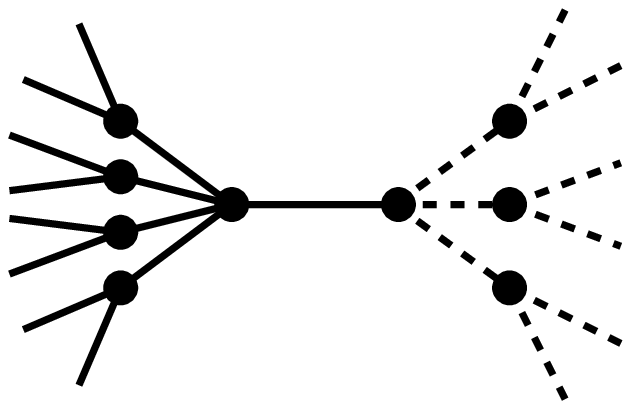} } \hfill 
  \subfloat[\label{blinker}]{
    \includegraphics[width=0.46\columnwidth]{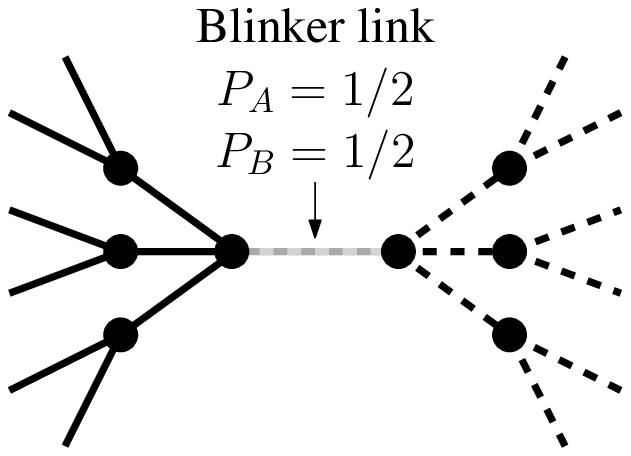} }
 \caption{Schematic illustration of disordered configurations found with a
 majority rule dynamics on link states with no rewiring ($p = 0$). a) Frozen
 disordered configuration. b) Dynamical trap based on a blinker link which keeps
 changing state forever with probability $1/2$.}
\label{disorderedConfs}
\end{figure}

The implementation of the majority rule that we use here is equivalent to the
zero-temperature Glauber dynamics \footnote{Different implementations are
possible, for example, by varying the probability to switch states in case of
tie.}, which has been extensively studied in the context of spin systems in
fixed networks and from a node states perspective. These studies show that, in
Erd\"os-R\'eny random networks, most realizations of the dynamics arrive to a
fully ordered, consensual state in a characteristic time which scales
logarithmically with system size \cite{Castellano2005, Baek2012}. However, a
very small number of runs (around a $0.02\%$ for $N=10^3$ and $\langle k \rangle
= 10$) end up in a disordered absorbing state, which can be frozen or
dynamically trapped \cite{Castellano2005}. The same disordered absorbing
configurations have also been found in \cite{Fernandez-Gracia2012} with a
prototype model of link-state majority rule dynamics. Nevertheless, the
probabilities are reversed: the frozen and dynamically trapped configurations
(see Fig.~\ref{disorderedConfs} for schematic examples) are the predominant
ones in link-based dynamics, while full order is only reached in very small and
highly connected networks.

In order to characterize the system at different times it is useful to consider
the \emph{density of nodal interfaces} $\rho$ as an order parameter
\cite{Fernandez-Gracia2012}, defined as the fraction of pairs of connected links
that are in different states. If $k_i$ is the degree of node $i$, and
$k_i^{A/B}$ is the number of $A/B$-links connected to node $i$ (with obviously
$k_i = k_i^A + k_i^B$), then $\rho$ is
calculated as:
\begin{equation}
 \rho = \frac{\sum_{i=1}^N k_i^{A} \, k_i^{B}}{\sum_{i=1}^N k_i (k_i - 1)/2}.
\end{equation}
The density $\rho$ is zero only when all connected links share the same state
and it reaches its maximum value of $1/2$ for a random distribution of states
(as it is the case in our initial condition), thus it is a measure of the local
order in the system. Note that complete order, $\rho = 0$, is achieved for both
connected consensual configurations, where all links are in the same state, and
configurations where the network is fragmented in a set of disconnected
components, each formed by links with the same state. In both cases complete
order is identified with absorbing configurations, where the system can no
longer evolve. In terms of the node-equivalent graph, the line-graph, the order
parameter $\rho$ becomes the \emph{density of active links}, i.e., the fraction
of links of the line-graph connecting nodes with different states.


\section{Final states}
\label{finalStates}

\begin{figure}[ht!]
  \subfloat[\label{runsRhoTimes}]{
    \includegraphics[width=0.345\columnwidth]{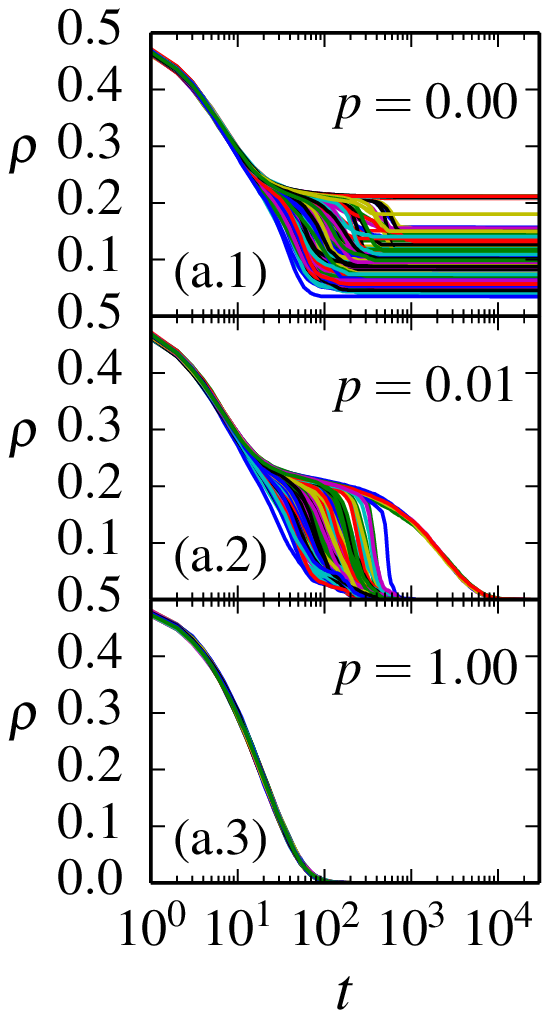}}
\hfill
  \subfloat[\label{avRhoTimes}]{
\includegraphics[width=0.625\columnwidth]{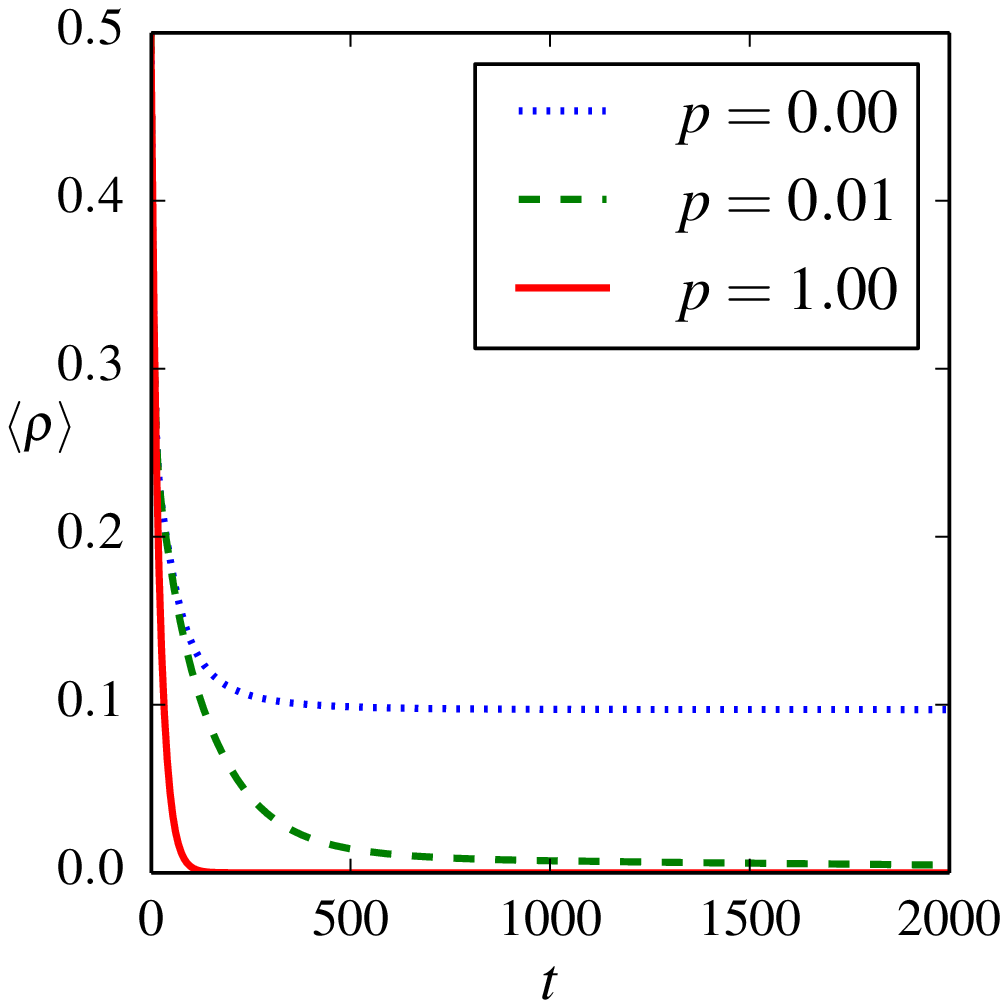} }
 \caption{(Color online) Behavior of the order parameter for a system with $N =
 2000$ and $\langle k \rangle = 10$. a) Density of nodal interfaces $\rho$ for
 $100$ individual realizations in linear-log scale. b) Average density of nodal
 interfaces $\langle \rho \rangle$ over $10000$ realizations. The time interval
 shown has been chosen for the sake of clarity; in reality, the runs for $p =
 0.01$ do not reach zero until $t \approx 30000$ while the ones for $p = 1.00$
 are zero from $t \approx 350$.} 
\label{rhoTimes}
\end{figure}

To explore how the coevolution of link states and network topology affects the
final state of the system we run numerical simulations of the dynamics
described above. The system evolves until the network reaches a final
configuration that strongly depends on the system size $N$ and the rewiring
probability $p$. The case $p=0$ corresponds to a static network situation,
analyzed in \cite{Fernandez-Gracia2012}. In this case, system sizes larger than
$N=500$ lead to disordered final states represented by network configurations
composed by several interconnected clusters of type $A$ and $B$ links. A link
that connects two clusters is either frozen, because it is in the local
majority, or switching ad infinitum between states $A$ and $B$ (``blinking''),
because it has the same number of neighboring links in each state. Therefore,
we refer to these as disordered configurations ($\rho > 0$) that are either
frozen or dynamically trapped, respectively (see Fig.~\ref{disorderedConfs}).
For $p>0$ the network always reaches an absorbing ordered configuration that can
be, either a one-component network with all links sharing the same state, or a
fragmented network consisting of two large disconnected components of size
similar to $N/2$ and in different states \footnote{A few disconnected nodes can
also be occasionally found.}. We remark that all links inside each component are
in the same state, thus the order parameter $\rho$ equals zero, as in the
non-fragmented case. The behavior of $\rho$ for different values of $p$ is shown
in Fig.~\ref{rhoTimes}, both as an average over different realizations
(\ref{avRhoTimes}) and as single trajectories (\ref{runsRhoTimes}). For $p=0$
almost every realization reaches a plateau or stationary value of $\rho > 0$
(see Fig.~\ref{runsRhoTimes}.1). For any $p>0$ every run reaches an ordered
absorbing state with $\rho = 0$ (see Figs.~\ref{runsRhoTimes}.2,
\ref{runsRhoTimes}.3). However, for small values of $p$ we observe a
distinction between two groups of realizations, one ordering much faster than
the other (see Fig.~\ref{runsRhoTimes}.2). These different time scales will be
discussed in section \ref{timeEvol}.

\subsection{Fragmentation transition in finite systems}
\label{fragTrans}

In order to explore how the network evolution affects the likelihood and the
properties of the two possible outcomes, one component or fragmentation in two
components, we study three relevant quantities. These are the probability $P_1$
that the final network is not fragmented, i.e, that it settles in one component,
the relative size $s_{\scriptscriptstyle L}$ of the largest network component
and the magnitude $\sigma_{s_L}$ of its associated fluctuations across different
realizations.

\begin{figure}[ht!]
 \centering \includegraphics[width=7cm, height=!]{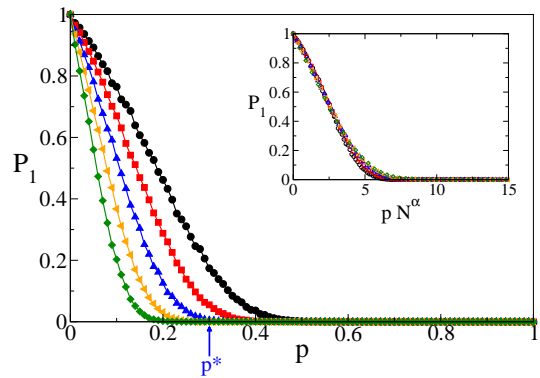}
 \caption{(Color online) Probability $P_1$ that the system ends in a single
 network component vs the rewiring probability $p$, for networks of mean degree
 $\mu=10$ and size $N=500$ (circles), $N=1000$ (squares), $N=2000$ (triangles
 up), $N=4000$ (triangles left) and $N=8000$ (diamonds).  $10000$ runs were used
 to estimate $P_1$, starting from an Erd\"os-R\'enyi network with random initial
 conditions. The limit of the region of bistability, $p^*$, is shown for the
 size $N=2000$. Note that $\forall p \ge p^*,\;P_1(p) < 1/N$. Inset: curves
 collapse when $p$ is rescaled by $N^{\alpha}$, with $\alpha = 0.42$.}
 \label{P1-p}
\end{figure}

In Fig.~\ref{P1-p} we show $P_1$ vs $p$, calculated as the fraction of
simulation runs that ended up in a single component. We observe that $P_1=1$
only for $p=0$, then it decreases continuously between $p=0$ and a
certain value $p=p^*$ and is always smaller than $1/N$ for $p \ge p^*$. This
defines three regimes regarding $p$: one point at $p=0$ where the system is
always connected, a region of bistability in $0<p<p^*$ where the system can
both stay connected in one piece or break into disconnected components, and a
fragmented region for $p \ge p^*$ where the network always splits apart.

\begin{figure}[ht!]
 \centering \includegraphics[width=7cm, height=!]{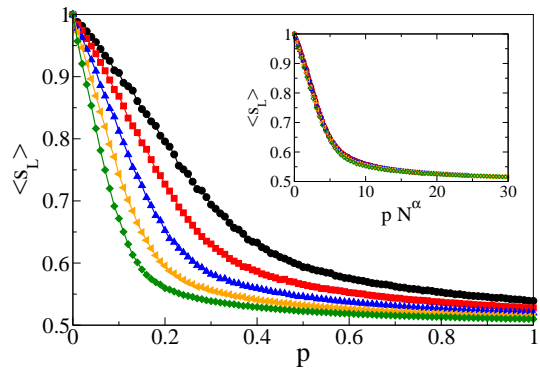}
 \caption{(Color online) Average relative size $\langle s_{\scriptscriptstyle L}
 \rangle$ of the largest network component  vs $p$, and for the same network
 sizes as in Fig.~\ref{P1-p}. $s_{\scriptscriptstyle L}$ is defined as the
 fraction of nodes included in the largest connected component. Inset: as in
 Fig.~\ref{P1-p}, $p$ is rescaled by $N^{\alpha}$, making the curves collapse to
 one.}
 \label{s-p}
\end{figure}

\begin{figure}[ht!]
 \centering \includegraphics[width=7cm, height=!]{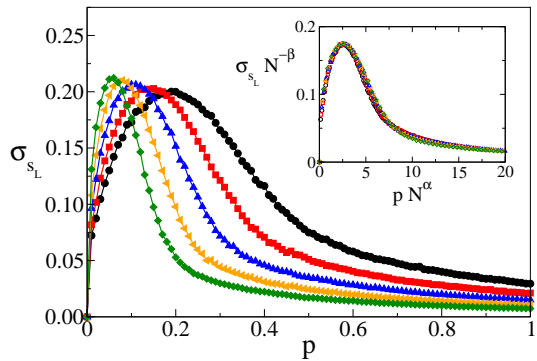}
 \caption{(Color online) Standard deviation $\sigma_{s_L}$ of the relative size
 $s_{\scriptscriptstyle L}$ of the largest network component for the same system
 sizes $N$ as in Fig.~\ref{P1-p}. $\sigma_{s_L}$ is a measure of the magnitude
 of the fluctuations in the final size of the largest network component across
 different realizations of the dynamics. Inset: collapse of all curves by
 rescaling $p$ by $N^{\alpha}$ and $\sigma_{s_L}$ by $N^{-\beta}$, with
 $\alpha=0.42$ and $\beta=0.022$.}
\label{sigma-p}
\end{figure}

\begin{figure}[ht!]
    \subfloat[\label{HistoSizes}]{
    \includegraphics[width=0.75\columnwidth]{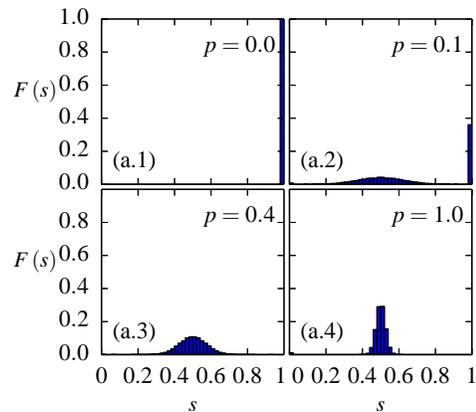} } \hfill 
  \subfloat[\label{Hysteresis}]{
    \includegraphics[width=0.75\columnwidth]{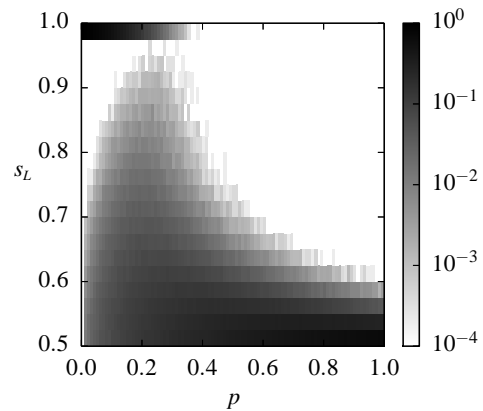} }
\caption{Relative sizes $s$ of network components for $N=2000$, $\langle k
\rangle = 10$ and $10000$ runs starting from random initial conditions. a)
Histogram of network relative sizes for four different probabilities of
rewiring $p$. b) Color-map of the fraction of runs ending in a given relative
size of the largest network component $s_{\scriptscriptstyle L}$. Note the
logarithmic color scale. White corresponds to no run ending in that relative
size.}
\label{FragSizes}
\end{figure}

This result is consistent with the behavior of the average value of
$s_{\scriptscriptstyle L}$ over many realizations (see Fig.~\ref{s-p}), which
decreases from $\langle s_{\scriptscriptstyle L} \rangle = 1$ for $p=0$ to
$\langle s_{\scriptscriptstyle L} \rangle \simeq 0.5$ for large $p$. As shown in
Fig.~\ref{sigma-p}, the standard deviation of $s_{\scriptscriptstyle L}$
($\sigma_{s_L}$) has its maximum at a value $p_{\mbox{\tiny max}}$ for which
$P_1$ is approximately $0.5$, that is, where fragmented and non-fragmented
realizations are equally probable. The peak in $\sigma_{s_L}$ indicates a broad
distribution of possible largest component sizes in that region and thus
$p_{\mbox{\tiny max}}$ can be used as a footprint of the transition point. This
broad distribution can also be seen in Fig.~\ref{Hysteresis}, where we present a
color-map of the fraction of runs that ended up in a given relative size
$s_{\scriptscriptstyle L}$ of the largest network component for a network of
$N=2000$ nodes. For the sake of clarity we also present in Fig.~\ref{HistoSizes}
histograms of network relative sizes $s$ (not only the largest) for four
different values of $p$. We note that the maximum of $\sigma_{s_L}$ occurs
around $p \approx 0.1$ (see Fig.~\ref{sigma-p}), which corresponds in the
color-map to a distribution of $s_{\scriptscriptstyle L}$ that has a peak at
$s_{\scriptscriptstyle L} = 1$ (one component) and a broad distribution
corresponding to fragmented cases with $0.5 \leq s_{\scriptscriptstyle L} \leq
0.875$. This division into fragmented and non-fragmented runs can also be
clearly observed in the histogram corresponding to $p=0.1$ (see
Fig.~\ref{HistoSizes}.2).

Interestingly, a common feature of $P_1(p)$, $s_{\scriptscriptstyle L}(p)$ and
$\sigma_s(p)$ curves is that they are shifted to smaller values of $p$ as the
system size $N$ increases, and thus the range of $p$ for which there is
bistability of fragmented and non-fragmented outcomes seems to vanish in the
thermodynamic limit, i.e., $p*$ tends to zero as size is increased. This
shifting behavior also points at the fact that the transition point
$p_{\mbox{\tiny max}}$ appears to tend to zero in the infinite size limit. A
dependence of the transition point with the system size, in a way that it tends
to zero in the infinite size limit, has been shown to be the case in several
opinion dynamics models \cite{Toral2007}. Such systems, as it is the case here,
do not display a typical phase transition in the thermodynamic limit with a
well defined critical point and its associated critical exponents, divergences
(in case of a continuous, second order phase transition) or discontinuities (in
case of a first order phase transition). However, for any finite system a
transition point can be clearly defined as separating two different behavioral
regimes.

To gain an insight about the $N \to \infty$ behavior, we perform a finite size
scaling analysis by assuming that $P_1$, $s_{\scriptscriptstyle L}$ and
$\sigma_{s_L}$ are functions of the variable $x \equiv p\,N^{\alpha}$:
\begin{equation}
\begin{aligned}
P_1(p,N) &= P_1(p\,N^{\alpha}),\\[5pt]
s_{\scriptscriptstyle L}(p,N) &= s_{\scriptscriptstyle L}(p\,N^{\alpha}),\\[5pt]
\sigma_{s_L}(p,N) &= N^{\beta} \sigma_{s_L}(p\,N^{\alpha}).
\end{aligned}
\end{equation}
The values of the exponents $\alpha$ and $\beta$ should be such that make the
curves for different sizes collapse into a single curve. Therefore, the location
of the peak in all $\sigma_{s_L}(p)$ curves of Fig.~\ref{sigma-p} should scale as
$p_{\mbox{\tiny max}} \sim N^{-\alpha}$. By fitting a power law function to the
plot $p_{\mbox{\tiny max}}$ vs $N$ we found $\alpha \simeq 0.42$ (not shown). In
the insets of Figs.~\ref{P1-p}, \ref{s-p} and \ref{sigma-p} we observe the
collapse for different network sizes when magnitudes are plotted versus the
rescaled variable $x$ (rescaling also the y-axis by $N^{-\beta}$ in the case of
$\sigma_{s_L}$). This scaling analysis shows that, in the thermodynamic limit,
the network would break apart for any finite value of $p>0$. This might be
related to the fact that when the system evolves under the majority rule alone,
it always gets trapped in disordered configurations (in the $N \to \infty$
limit). Then, it seems that even a very small rewiring rate is enough to remove
the system from traps, but at the cost of breaking the network apart. However,
as we will show in the next section, the time needed for the fragmentation to
occur diverges with system size. A deeper understanding of this phenomenon can
be achieved by studying stochastic trajectories of single realizations.


\section{Time evolution}
\label{timeEvol}

\begin{figure}[ht!]
 \centering \includegraphics[width=\columnwidth,height=!]{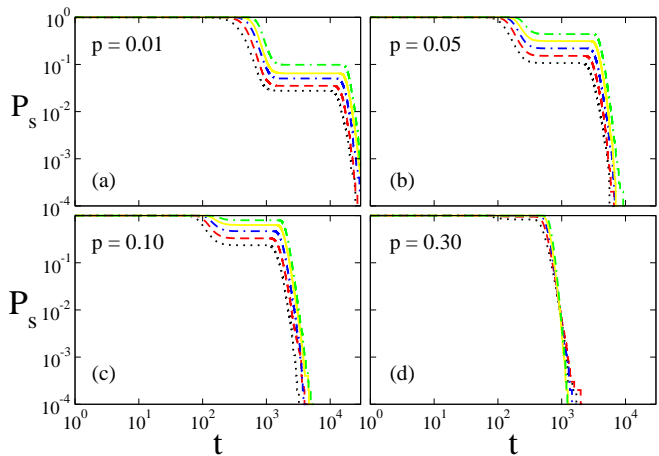}
 \caption{(Color online) Time evolution of the survival probability $P_s$ for
 different values of $p$ and networks of size $N=500$, $1000$, $2000$, $4000$
 and $8000$ (curves from bottom to top and, respectively, dotted, dashed,
 dash-dotted, solid and dash-dash-dotted). Averages are over $10^4$ independent
 runs.}
 \label{Ps-t}
\end{figure}

We are interested in quantifying the evolution of the system towards the final
states described above. In Fig.~\ref{Ps-t} we plot the survival probability
$P_s(t)$, i.e, the probability that a realization did not reach the ordered
state ($\rho=0$) up to time $t$.

When $p=0$ we have $P_s=1$ for all times, meaning that all realizations (except
for a few runs with the smallest size $N=500$, as reported in
\cite{Fernandez-Gracia2012}) fall into a disordered configuration characterized
by a constant value of $\rho>0$, as we shall discussed in detail in the next
section. For $p=0.01$, $p=0.05$ and $p=0.10$ [Figs.~\ref{Ps-t}(a), \ref{Ps-t}(b)
and \ref{Ps-t}(c), respectively] we observe that $P_s$ experiences two decays at
very different time scales, revealing the existence of two different ordering
mechanisms. As we will explain, the first decay from $P_s =1$ to a plateau
corresponds to the ordering of non-fragmented realizations, while the second
decay from the plateau to zero is due to the ordering of fragmented runs. Take,
for instance, $p=0.01$ and $N=8000$. We observe in Fig.~\ref{P1-p} that the
fraction of runs ending in one component is $P_1 \simeq 0.9$. We interpret that
it is the arrival of this $90\%$ of runs to a one-component absorbing state with
$\rho=0$ which produces the first decay of the survival probability to $P_s
\simeq 0.1$ around a time $t \simeq 10^3$, as can be observed in
Fig.~\ref{Ps-t}(a). The remaining fraction $P_s \simeq 0.1$ that survive lead to
the plateau that lasts up to the second decay around $t \simeq 10^4$, when they
arrive to a fragmented absorbing state again with $\rho=0$. Note also that both
decay times decrease for increasing $p$, while the height of the plateau rises
($P_1$ increases). In the $p=0.30$ case [Fig.~\ref{Ps-t}(d)] the first decay of
$P_s$ is only observed for small systems, since for larger ones most
realizations end up with a fragmented network (see Fig.\ref{P1-p}). This picture
also holds for larger values of $p$.


\subsection{Description of trajectories in phase space}
\label{trajectories}

\begin{figure}
 \subfloat[$p = 0.00$\label{Traj-0.00}]{
   \includegraphics[width=0.48\columnwidth]{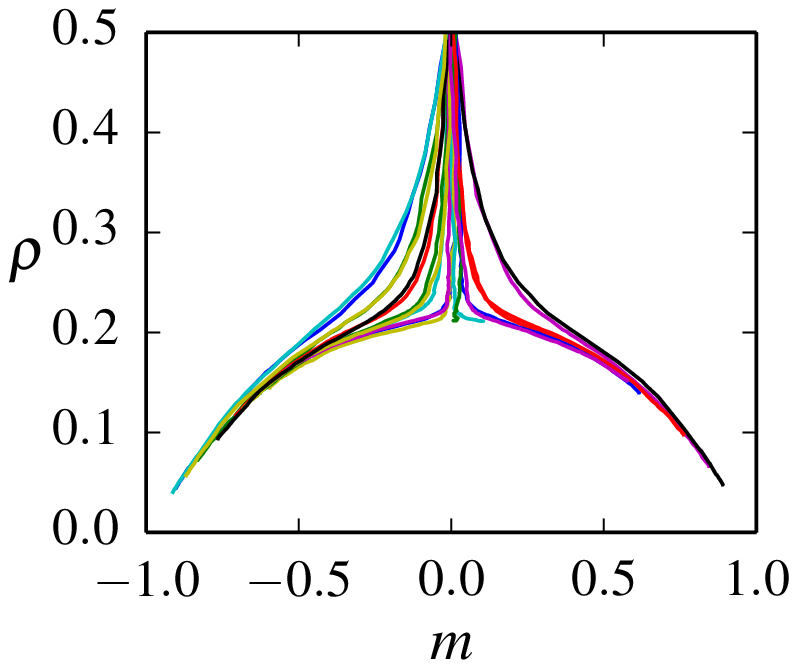}
 }
 \hfill
 \subfloat[$p = 0.01$\label{Traj-0.01}]{
   \includegraphics[width=0.48\columnwidth]{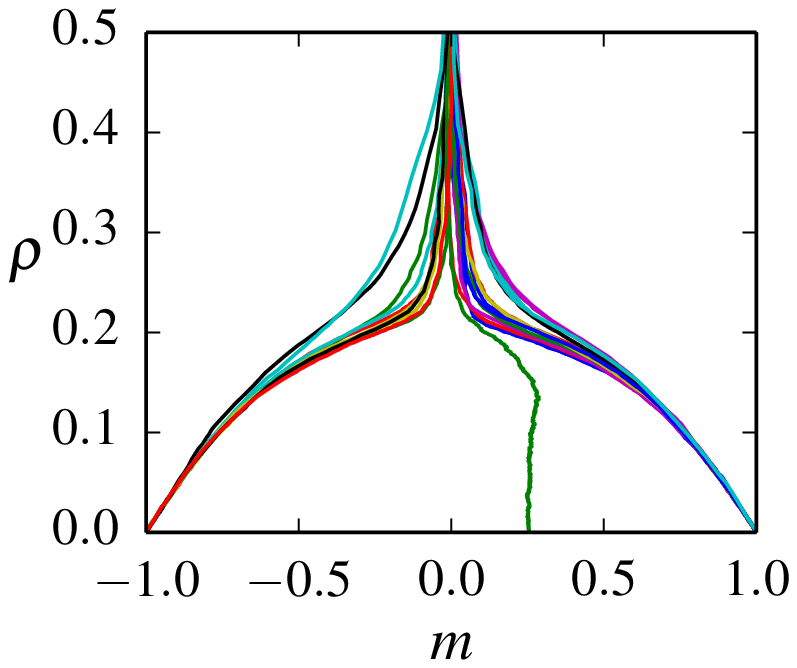}
 }
 \hfill
 \subfloat[$p = 0.10$\label{Traj-0.10}]{
   \includegraphics[width=0.48\columnwidth]{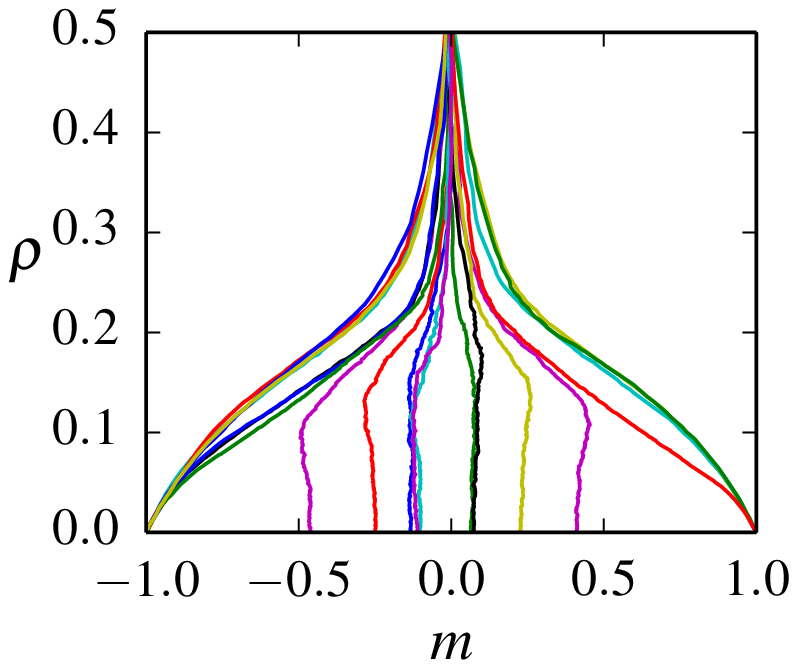}
 }
 \hfill
 \subfloat[$p = 0.50$\label{Traj-0.50}]{
   \includegraphics[width=0.48\columnwidth]{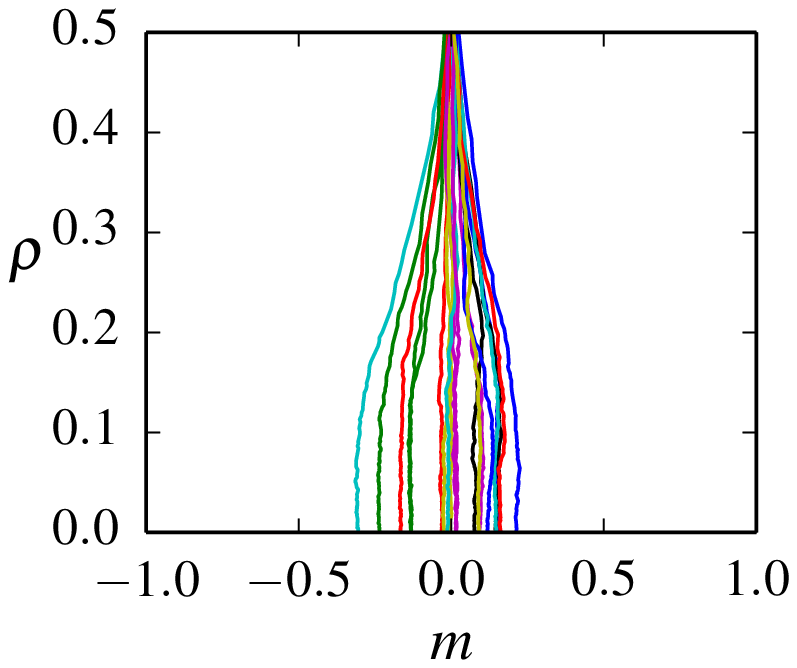}
 }
 \caption{(Color online) Typical trajectories of the system on the $(m,\rho)$
 space for a  network of $N=2000$ nodes and different values of the rewiring
 probability $p$.}
\label{Traj}
\end{figure}

In order to gain an insight about the fragmentation phenomenon, we investigate
in this section individual trajectories of the system on the $m-\rho$ plane,
where $m$ is the link magnetization \cite{Vazquez2008a,Vazquez2008b}, the
difference between the fractions of $A$ and $B$ links,
\begin{equation}
  m = \frac{\sum_{i=1}^N \left( k_i^{A} - k_i^{B} \right)}
  {\sum_{i=1}^N k_i}.
\end{equation}
In Fig.~\ref{Traj} we display typical trajectories of the system for a network
of $N=2000$ nodes and values of the rewiring probability $p=0, 0.01, 0.1$ and
$0.5$. Trajectories start at $(m,\rho) \simeq (0,0.5)$, corresponding to random
initial conditions. Points $(1,0)$ and $(-1,0)$ represent $A$ and $B$
one-component consensual configurations, while the absorbing line $\rho=0$ with
$|m|<1$ corresponds to a fragmented network.

In the $p=0$ case (Fig.~\ref{Traj-0.00}), we observe that realizations undergo a
fast initial ordering in which associated trajectories go from $\rho \simeq 0.5$
to $\rho \simeq 0.2$  (with some small changes in $m$) in approximately $25$
Monte Carlo steps. This corresponds to the fast formation of two giant
(connected) domains of opposite states due to the majority rule dynamics, as has
been reported in previous works \cite{Castellano2006}. Afterwards trajectories
enter in a common curve which, as in other cases \cite{Vazquez2008a}, can be fit
by a parabola and where the ordering process is accompanied by a change in
magnetization. In our case the parabola takes the approximate form $\rho \simeq
0.2(1-m^2)$ and the system evolves following a direct path towards $|m| = 1$,
due to the fact that $\rho$ cannot increase in a majority rule update. This
corresponds to the largest domain progressively invading the other. However, the
ordering stops abruptly when the system falls to a topologically trapped state
with $\rho >0$, preventing it from arriving to the one-component ordered $A$ or
$B$ states, $(1,0)$ or $(-1,0)$ points, respectively.

For $p = 0.01$ (Fig.~\ref{Traj-0.01}) most runs finally arrive to the
one-component ordered state, by means of the rewiring mechanism that helps the
system escape from frozen or dynamical traps. As mentioned before, even a small
rewiring rate is able to unlock frustrated links, allowing the system to keep
evolving towards one-component order ($|m|=1$, $\rho=0$). Nevertheless, there
are some runs that escape from the parabola and follow a nearly vertical
downward trajectory (line ending at $\rho=0$ and $m \simeq 0.25$), even if they
are initially attracted towards $|m| = 1$. These runs are trapped around a given
value of $m$ and experience a relaxation that decreases $\rho$ very slowly while
keeping $m$ almost constant. It seems that in these realizations some rewiring
events trigger only a few successful majority rule updates that are not enough
to completely order the system in a one-component network. This corresponds to
the process of fragmentation of the network in two components with different
states. For larger rewiring rates more runs end up fragmenting in two components 
(see Fig.~\ref{Traj-0.10}), until for large enough $p$ no run is able to follow
the parabola (see Fig.~\ref{Traj-0.50}), leading to only fragmented final
states.


\subsection{Mean-field approach}
\label{meanField}

As explained in the last section and shown in Fig.~\ref{rhoTimes}, $\langle \rho
\rangle$ undergoes a first fast decay in a short time scale corresponding to the
contribution of non-fragmented realizations, and then a second much slower decay
that corresponds to fragmented realizations. Therefore, bearing in mind that
much of the time evolution of $\langle \rho \rangle$ is controlled by the second
very slow dynamics of fragmenting realizations, we develop in this section an
analytical approach for this second regime. We assume that the system starts at
$t=0$ from a trapped configuration (see Fig.~\ref{disorderedConfs}), which
consists of two network components of similar size $N/2$ interconnected by
frustrated links. These are links with the same state as the majority of their
neighboring links, thus they cannot change state (see Fig.~\ref{frozen}), or
links with equal number of neighbors in each state, thus they keep flipping state
from $A$ to $B$ and vice versa (blinkers, see Fig.~\ref{blinker}). To estimate
how the density of frustrated links $\beta$ varies with time, we now describe the
events and their associated probabilities that lead to a change in $\beta$. In a
single time step of interval $dt=1/N$, a frustrated link is chosen with
probability $\beta$. Then, with probability $p/2$ the end of the link connected
to the minority is randomly chosen and rewired to another random node in the
network. Finally, this end lands on the component that holds the link's state
with probability $1/2$. After the rewiring this link is no longer connecting
components, thus the number of frustrated links is reduced by $1$, leading to a
change $\Delta \beta = - 2/\mu N$ (with $\mu \equiv \langle k \rangle$, as
above). Assembling all these factors, the average density of frustrated links
evolves according to
\begin{eqnarray}
\frac{d \beta(t)}{dt} = - \frac{p}{2 \mu} \beta(t),
\end{eqnarray}
with solution
\begin{eqnarray}
\beta(t) = \beta_0 \, e^{-\frac{p}{2\mu} t},
\end{eqnarray}
where $\beta_0$ is the initial density of frustrated links. Given that, on
average, each frustrated link accounts for the existence of $\mu-1$ nodal
interfaces, $\rho$ is proportional to $\beta$, and therefore we expect that the
average density of interfaces decays as
\begin{eqnarray}
\rho(t) \sim e^{-\frac{p}{2\mu} t}.
\label{rho-ave-t}
\end{eqnarray}

\begin{figure}[ht!]
 \centering \includegraphics[width=7cm,height=!]{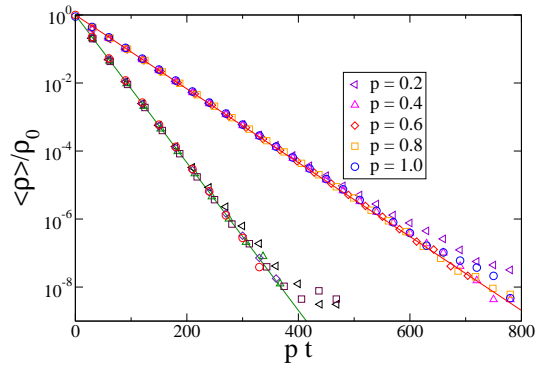}
 \caption{(Color online) Time evolution of the average density of nodal
 interfaces $\langle \rho \rangle$ on a linear-log scale, for values of the
 rewiring probability $p$ as indicated in the box. Symbols at the top correspond
 to simulations on a network of $N=4000$ nodes and mean degree $\mu=20$, while
 bottom symbols are for a network of size $N=8000$ and mean degree $\mu=10$.
 Time is rescaled by $p$ and $\langle \rho \rangle$ is normalized by its initial
 value to make the data collapse. Solid lines are the analytical approximations
 from Eq.~\eqref{rho-ave-t}.} 
\label{rho-t-20-40}
\end{figure}

In Fig.~\ref{rho-t-20-40} we show $\langle \rho \rangle$ vs time obtained from
numerical simulations for various values of $p$ (symbols) and two different
networks, one of size $N=8000$ and $\mu = 10$ and the other with $N=4000$ nodes
and $\mu = 20$. We observe that the expression~\eqref{rho-ave-t} (solid lines)
captures the behavior of $\langle \rho \rangle$ for most values of $p$ and has
the correct scaling with $\mu$. The data for $p=0.2$ deviates from the pure
exponential decay at long times, probably because the analytical approximation
works better for large $p$, where the rewiring process seems to dominate the
dynamics.


\subsection{Convergence times}
\label{convTimes}

\begin{figure}[ht]
 \centering \includegraphics[width=8cm, height=!]{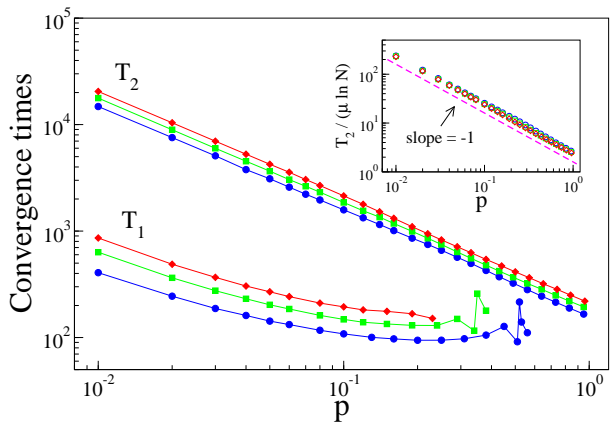}
 \caption{(Color online) Mean time to reach the fragmented and non-fragmented
 final states $T_1$ and $T_2$, respectively, vs the rewiring probability $p$,
 for networks of size $N=500$ (circles), $N=2000$ (squares) and $N=8000$
 (diamonds), and mean degree $\mu=10$.  The inset shows the scaling of $T_2$ as
 described by Eq.~\eqref{T2-p}.}
 \label{tau-p}
\end{figure}

Another quantity that is worth studying in this system is the time to reach the
final state, or convergence time, given that it complements our previous
analysis of the two ordering dynamics, majority rule and rewiring. In
Fig.~\ref{tau-p} we show the mean time of convergence to the final ordered
state for non-fragmented and fragmented runs $T_1$ and $T_2$, respectively,
versus the rewiring probability $p$ \footnote{Here the subindices $1$ and $2$
refer to one and two components, even though fragmented runs may also have a
few disconnected nodes}. Results are shown for three different system sizes. We
observe that $T_2$ is about ten times larger than $T_1$ for all values of $p$.
This confirms the dynamical picture that we discussed in the previous sections.
There is a first fraction of runs in which the majority rule dynamics plays a
leading role constantly ordering the system until it reaches one-component full
order in a short time scale $T_1$. But there is also a second fraction which
fall into particular topological traps that prevent the system to keep ordering,
and then the rewiring process slowly leads to the fragmentation of the network
in a much longer time scale $T_2$. Interestingly, rewiring always works as a
perturbation that frees the system whenever it gets trapped, but it seems that
in the first type of runs perturbations trigger cascades of ordering updates
which are large enough to completely order the network before it breaks apart.

\begin{figure}[ht!]
 \centering \includegraphics[width=7cm, height=!]{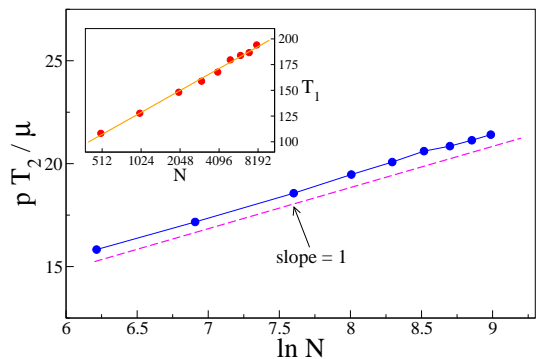}
 \caption{(Color online) Convergence times $T_1$ and $T_2$ vs system size $N$
 for $p=0.1$ and $\mu=10$.  Main: y and x-axis were rescaled according to
 Eq.~\eqref{T2-p}. Inset: data is shown on a log-linear scale. The solid line is
 the best fit $T_1 = 30.7 \, \ln N - 84.5$.}
 \label{tau-N}
\end{figure}

An approximate expression for $T_2$ can be obtained by considering the
relaxation to the fragmented state given by Eq.~\eqref{rho-ave-t}, where the
mean number of nodal interfaces decreases to zero. The network breaks in two
components when the fraction of frustrated links holding both components
together becomes smaller than $2/\mu N$, or $\rho \sim 1/N$, since $\rho$ is
proportional to $\beta$, as we mentioned before. Then, we can write $1/N \sim
\exp(-p \, T_2/2 \mu)$, from where 
\begin{equation}
T_2 \sim \frac{\mu}{p} \ln N. 
\label{T2-p}
\end{equation}
The inset of Fig.~\ref{tau-p} shows that the approximate expression \eqref{T2-p}
captures the right scaling of $T_2$ with $p$ and $N$. In Fig.~\ref{tau-N} we
check the dependence of $T_1$ and $T_2$ with the system size $N$. The y-axis of
the main plot showing $T_2$ was rescaled according to Eq.~\eqref{T2-p}. The
inset shows that $T_1$ also scales as $\ln N$.

\begin{figure}[ht!]
 \centering \includegraphics[width=7cm, height=!]{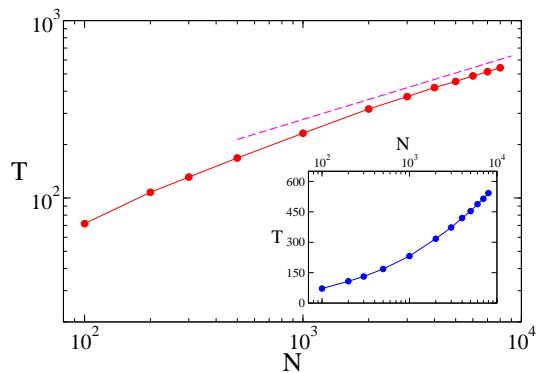}
 \caption{(Color online) Average time to reach an absorbing disordered state $T$
 vs systems size $N$ on a double logarithmic scale, for a static network ($p=0$).
 The dashed line has slope $0.375$. The log-linear scale in the inset shows that
 $T$ grows faster than $\ln N$.}
 \label{tau-N-0}
\end{figure}

As Fig.~\ref{tau-p} shows, both $T_1$ and $T_2$ decay as $1/p$ in the low $p$
limit. This is because when $p$ is very small we can picture a typical evolution
of the system as a series of alternating pinning and depinning processes. That
is, initially a series of majority rule updates take place, which partially
order the system until it reaches a frustrated configuration. Then the system
stays trapped there for a time of order $1/p$ until a successful rewiring event
unlocks it. This is followed by another avalanche of majority rule updates that
ends on the next trapped state. This process is repeated until a final absorbing
ordered configuration is reached. Given that the mean time interval between two
avalanches scales as $1/p$, the convergence time to any final state should scale
as $1/p$ (see Fig.~\ref{tau-p}). This implies that $T_1$ and $T_2$ diverge as $p
\to 0$. However, when $p$ is strictly zero the system is absorbed in a disordered
configuration, which can be frozen or dynamically trapped, and so the convergence
time is finite. The $p=0$ case also differs from the $p>0$ case in the fact that
convergence times to the absorbing disordered configurations seem to scale as $T
\sim N^{0.375}$ (see Fig.~\ref{tau-N-0}), instead of $\ln N$. 

\vspace{0.5cm}


\section{Summary and Conclusions}
\label{conclusions}

We have studied a model that explores the majority rule link dynamics on a
coevolving network, where links in the local minority are rewired at random. On
topologically static ($p=0$) large networks, the ordering process induced by the
majority rule stops before a completely ordered state is reached with all links
in the same state (the only possibility with no rewiring), because the system
falls into trapped disordered configurations. When the rewiring is switched on
($p>0$), the system is able to escape from these trapped configurations and
reach an ordered absorbing state that can be either a one-component network with
all links in the same state or a fragmented network with two opposed states
disconnected components. The former output is more likely when the rewiring rate
is low or networks are small, while the latter output becomes more and more
common as the rewiring rate increases or networks get larger, and it is the only
possible result for large rewiring rates or in the limit of very large networks.
For any finite size network, a range of values of the rewiring probability $p$
can be found for which there is bistability between both possible outcomes. In
the very large size limit, however, the bistability region progressively vanishes
and thus even very small amounts of rewiring make the network break apart.

By studying the trajectories of the system in the $m-\rho$ space we were able to
identify two types of evolutions, which provides an insight about the mechanism
of fragmentation. For no rewiring, all trajectories fall into an attractive path
with a parabolic envelope that ends in a point corresponding to a one-component
ordered configuration. However, these trajectories stop before reaching that
point, indicating that the system is trapped in a disordered configuration. For
low rewiring, most trajectories quickly move along the parabola until they hit
the one-component ordered absorbing point. This complete ordering process is
mainly driven by majority rule updates, and happens in a quite short time scale.
For high rewiring a new scenario appears. Most trajectories quickly stop at some
point in the parabola, and then slowly follow a nearly vertical path that ends
in the absorbing line $\rho=0$ with $|m|<1$, corresponding to a fragmented
network. This second fragmentation process takes a much longer time than the
initial ordering process, and controls the total convergence time to the final
state.

Our results show that the frozen and dynamically trapped disordered
configurations promoted by the link-based majority rule dynamics are not robust
against topological perturbations in the form of a rewiring, since the
continuous relinking updates are able to remove the system from the topological
traps. However, if instead of topological perturbations we consider
perturbations on the state dynamics in the form of a temperature, as in a
Glauber dynamics with a non-zero temperature, we find that the frozen and
dynamically trapped configurations appear to be robust for small noise
intensities \footnote{A. Carro, M. San Miguel, R. Toral, unpublished.}. Indeed,
even if any finite system with finite temperature perturbations is expected to
order by finite-size fluctuations, the ordering times become so large even for
small systems that, in practice, one can consider them as permanently trapped
in a disordered configuration.

By adopting a link-state perspective, our research contributes to the
understanding of complex phenomena emerging from the coupling of diffusive
processes with time varying networks. However, both reference
\cite{Fernandez-Gracia2012} and this paper are limited to states defined on the
links. A natural step further would be to consider mixed dynamics, with states
defined both on the nodes and on the links and a certain coupling between them.
Continuing with the language competition example used above, the node dynamics
would correspond to the evolution of language competence or preference, while
the dynamics on the links would mimic the evolution of language use. Work along
these lines is in progress.

\begin{acknowledgments}
We are particularly grateful to Juan Fern\'andez Gracia for helpful suggestions
during the early stages of this work. We acknowledge financial support by the EU
(FEDER) and the Spanish MINECO under Grant INTENSE@COSYP (FIS2012-30634) and by the
EU Commission through the project LASAGNE (FP7-ICT-318132).
\end{acknowledgments}

\bibliography{LinkStates}

\end{document}